\documentclass[10pt]{iopart}

\usepackage{graphicx}
\usepackage{dcolumn}
\usepackage{bm}
\usepackage{color}

\begin{document}

\title[MHD burst]{Simulation of the TAEs saturation phase in Large Helical Device device: MHD burst}


\author{J. Varela}
\ead{jvrodrig@fis.uc3m.es}
\address{Universidad Carlos III de Madrid, 28911 Leganes, Madrid, Spain}
\address{National Institute for Fusion Science, National Institute of Natural Science, Toki, 509-5292, Japan}
\address{Oak Ridge National Laboratory, Oak Ridge, Tennessee 37831-8071, USA}
\author{D. A. Spong}
\address{Oak Ridge National Laboratory, Oak Ridge, Tennessee 37831-8071, USA}
\author{Y. Todo}
\address{National Institute for Fusion Science, National Institute of Natural Science, Toki, 509-5292, Japan}
\author{L. Garcia}
\address{Universidad Carlos III de Madrid, 28911 Leganes, Madrid, Spain}
\author{Y. Ghai}
\address{Oak Ridge National Laboratory, Oak Ridge, Tennessee 37831-8071, USA}
\author{J. Ortiz}
\address{Universidad Carlos III de Madrid, 28911 Leganes, Madrid, Spain}
\author{R. Seki}
\address{National Institute for Fusion Science, National Institute of Natural Science, Toki, 509-5292, Japan}

\date{\today}

\begin{abstract}
The aim of the present study is to analyze the saturation regime of the Toroidal Alfven Eigenmodes (TAE) in the LHD plasma, particularly the MHD burst. The linear and nonlinear evolution of the TAEs are simulated by the FAR3d code that uses a reduced MHD model for the thermal plasma coupled with a gyrofluid model for the energetic particles (EP) species. The linear simulations indicate the overlapping of $1/2-1/1$, $2/3-2/4$ and $3/5-3/6$ TAEs in the inner-middle plasma region and frequency range of $45-75$ kHz, triggered by EPs with an energy of $T_{f} = 45$ keV and EP $\beta = 0.022 $. The nonlinear simulations show that $2/3-2/4$ and $3/4-3/5$ TAEs are further destabilized due to the energy transfer from $1/1-1/2$ TAE, leading to a broad TAEs radial overlapping and the MHD burst triggering. The energy of $1/1-1/2$ TAE is also nonlinearly transferred to the thermal plasma destabilizing the $0/0$ and $0/1$ modes, inducing the generation of shear flows and zonal currents as well as large deformations in the thermal pressure and EP density radial profiles. The nonlinear simulation reproduces the same succession of instabilities and the same frequency range with respect to the experiment. The instability propagates outward during the bursting phase, showing a large decrease of the EP density profile between the middle-outer plasma, pointing out the loss of part of the EP population that explains the decrease of the plasma heating efficiency observed during the MHD burst.

\end{abstract}

%
%
%
%
%

\pacs{52.35.Py, 52.55.Hc, 52.55.Tn, 52.65.Kj}

\vspace{2pc}
\noindent{\it Keywords}: Stellarator, LHD, EIC, MHD, AE, energetic particles

\maketitle

\ioptwocol

\section{Introduction \label{sec:introduction}}

The plasma destabilization by energetic particles (EP) injected via neutral beam injectors (NBI) is routinely observed in Large Helical Device (LHD) discharges \cite{1,2,3,4,5,6,7,8,9}. EP driven instabilities enhance the EP transport leading to EP losses before thermalization, reducing the LHD performance due to an inefficient plasma heating \cite{10,11,12,13,14,15,16,17,18,19}. 

EP driven instabilities as the Alfv\'en eigenmodes (AE) and the energetic particle modes (EPM) are destabilized if there is a resonance between the EP drift, bounce or transit frequencies and the AE / EPM frequency \cite{20}. The AEs are driven in the spectral gaps of the shear Alfv\' en continua \cite{21,22} although the EPMs are triggered in the shear Alfven continua if the continuum damping is not strong enough to stabilize them \cite{23,24,25}. Different families of AEs exist, destabilized in frequency gaps associated with periodic variations of the Alfv\' en speed. Present study is dedicated to the analysis of the toroidicity induced AE (TAE), coupling $m$ with $m+1$ modes ($n$ is the toroidal mode and $m$ the poloidal mode)\cite{26,27}.

LHD plasma is heated using NBI parallel and perpendicular lines generating EP with initial energies of $180$ and $32$ keV, respectively. The EP resonance is particularly large in LHD operation scenarios with low magnetic field and density, leading to the destabilization of AEs if the plasma is strongly heated by the tangential NBIs. Particularly, high $\beta$ discharges in LHD inward shifted configuration with low magnetic fields ($\approx 0.5$ T) show bursting MHD activity, leading to an enhancement of the EP fluxes to the tangential neutral particle analyzer (NPA) (see fig 9 of reference \cite{28}). Such MHD activity are identified in the Mirnov coils signal as AE bursting activity in the frequency range of $40-80$ kHz, combined with an enhanced detection of high energy EP around $135$ keV. The analysis of the continuum gap structure and EP distribution function indicates the destabilization of $n/m = 1/1-1/2$ TAEs in the middle-outer plasma region (see fig 10 of reference \cite{28}). In addition, the frequency spectra shows multiple signals pointing out the destabilization of several AEs. TAE bursts were originally observed in Hydrogen plasma heated by Hydrogen NBIs, although recent experiments showed the destabilization of TAE burst in Deuterium plasma \cite{29}.

Present study is dedicated to reproduce the TAE saturation and MHD burst destabilization in Hydrogen plasma. On that aim, a set of linear and nonlinear simulations are performed using the FAR3d gyro-fluid code \cite{30,31,32,33}. The FAR3d code solves the reduced nonlinear resistive MHD equations coupled with the moment equations of the energetic ion density and parallel velocity \cite{34,35,36}. The Landau damping/growth is included by Landau closure relations, reproducing the linear wave-particle resonance effects on six field variables that evolve from a three dimensional equilibria generated by the VMEC code \cite{37}. 

This paper is organized as follows. The numerical scheme and equilibrium properties are described in section \ref{sec:model}. The linear stability of the AEs is studied in section \ref{sec:lin}. The analysis of the TAE saturation and MHD burst destabilization is performed in section \ref{sec:nlin}. Next, the conclusions of this paper are presented in section \ref{sec:conclusions}.

\section{Numerical scheme \label{sec:model}}

The numerical model consists in a reduced set of equations for high-aspect ratio configurations and moderate $\beta$-values (of the order of the inverse aspect ratio), that retains the toroidal angle dependency based upon an exact three-dimensional VMEC equilibrium (closed nested flux surfaces assumed) \cite{38}. The destabilizing effect induced by the EP population is added towards moments of the gyro-kinetic equation, particularly the EP density ($n_{f}$) and the EP velocity parallel to the magnetic field lines ($v_{||f}$). The model calibration requires the derivation of Landau closure coefficients obtained from gyro-kinetic simulations, matching the analytic TAE growth rates of the two-pole approximation of the plasma dispersion function, consistent with a Lorentzian energy distribution function for the energetic particles. The lowest order Lorentzian can be matched either to a Maxwellian or to a slowing-down distribution by choosing an equivalent average energy. Please see the references \cite{39,40} for further details of the model equations and numerical scheme.

FAR3d code was already validated with respect to gyro-kinetic and hybrid codes in dedicated benchmarking studies \cite{41}. In addition, the linear simulation performed by the code show a reasonable agreement with respect to observational data, reproducing the AE stability in LHD \cite{42,43,44,45}, DIII-D \cite{46,47,48,49,50}, TJ-II \cite{33,51,52} and Heliotron J \cite{53,54} plasma. In addition, the nonlinear version of the code succeeded analyzing the sawtooth-like, internal collapse events and EIC burst observed in LHD plasma \cite{55,56,57,58,39} and the AE saturation in DIII-D plasma \cite{59}.

\subsection{Equilibrium properties}

\begin{figure}[h!]
\centering
\includegraphics[width=0.45\textwidth]{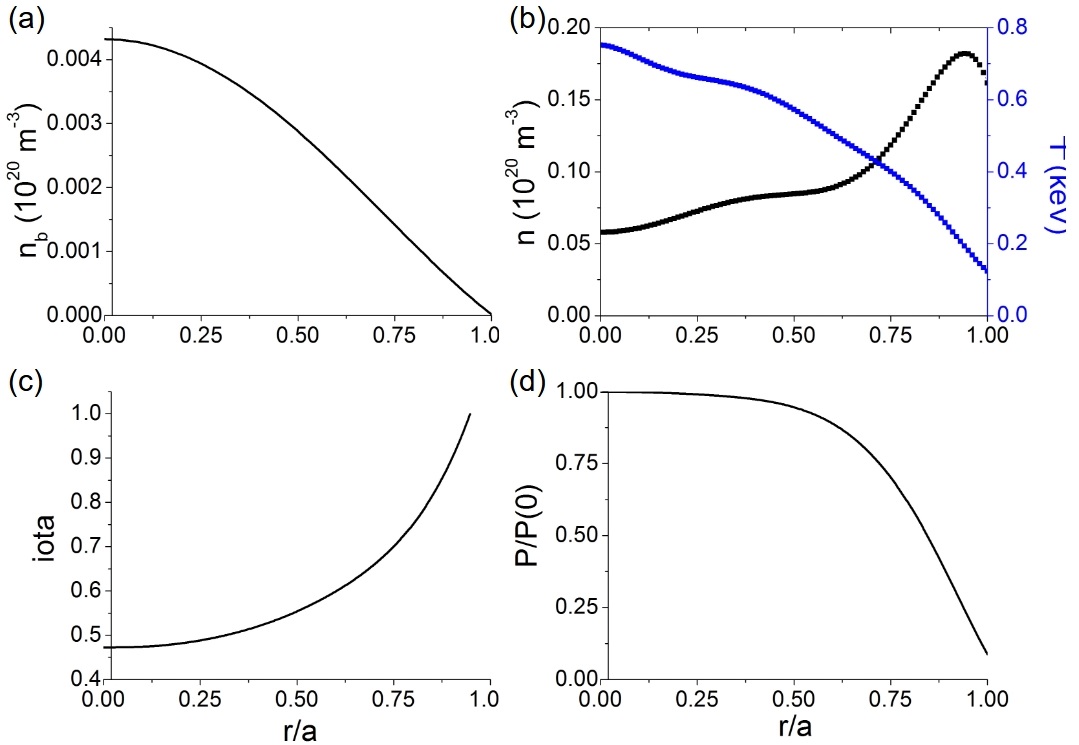}
\caption{(a) EP density profile, (b) thermal plasma density and temperature, (c) iota profile, (d) normalized pressure profile.}\label{FIG:1}
\end{figure}

A fixed boundary equilibrium of the LHD discharge $47645$ is calculated using the VMEC code \cite{37}. The shot $47645$ is a low magnetic field and bulk plasma density discharge with $R_{axis} = 3.76$ m and a magnetic field intensity at the magnetic axis of $0.619$ T. The electron density and temperature profiles were reconstructed by Thomson scattering data and electron cyclotron emission. Figure~\ref{FIG:1} indicates the model profiles for the thermal plasma and EP.

The birth energy of the particles injected by the tangential NBI is $T_{f}(0) = 180$ keV. The reference model assumes the nominal energy $T_{f}(0) = 45$ keV ($v_{th,f0} = 2.08 \cdot 10^{6}$ m/s) and an EP $\beta = 0.022$ ($\beta_{f}$), representing the EP population that causes the largest destabilizing effect, consistent with the analysis performed using MEGA code on the same topic \cite{60,61}.

\subsection{Simulation parameters}

Table~\ref{Table:1} indicates the dynamic and equilibrium toroidal ($n$) and poloidal ($m$) modes included in the linear and nonlinear simulations. The linear simulations are limited to $n=1$ to $5$ toroidal modes. The mode selection includes all the resonant modes between the magnetic axis and the plasma periphery. The number of point of the radial grid is $400$ for the nonlinear simulations and $1000$ for the linear simulations. 

\begin{table}[h]
\centering
\begin{tabular}{c | c }
\hline
$n$ & $m$  \\ \hline
$0$ & $[0 , 9]$ \\
$1$ & $[1 , 5]$ \\
$2$ & $[1 , 8]$ \\
$3$ & $[2 , 10]$ \\
$4$ & $[3 , 12]$ \\
$5$ & $[5 , 14]$ \\
$6$ & $[6 , 16]$ \\
$7$ & $[8 , 17]$ \\
$8$ & $[10 , 19]$ \\
$9$ & $[12 , 20]$ \\
$10$ & $[12 , 22]$ \\
\end{tabular}
\caption{Dynamic and equilibrium toroidal (n) and poloidal (m) modes in the linear and nonlinear simulations.} \label{Table:1}
\end{table}

Both parities must be included for all the dynamic variables because the moments of the gyro-kinetic equation breaks the MHD symmetry. The Fourier decomposition follows the next convention in the code, for example for the case of the pressure: $n > 0$ corresponds to $cos(m\theta + n\zeta)$ and $n<0$ corresponds to $sin(m\theta + n\zeta)$. The magnetic Lundquist number is assumed $S=10^7$, consistent with the S value at the middle plasma region of LHD plasma. The nonlinear simulations include a diffusivity of $D_{i} = 10^{-5}$ for each variable (normalized to the Alfven time, $\tau_{A0}$, and the minor radius) and the EP Finite Larmor radius damping effects from $t > 400 \tau_{A0}$ (once the simulation is in the nonlinear phase). EP FLR effects are not included in the linear phase to speed up the simulation, enhancing the AE destabilization to reach faster the nonlinear phase. In the appendix B there is a description of the Larmor radius effect on $n=1$ to $5$ AEs growth rate and frequency for simulations including the EP FLR damping.

The representation of the eigenfunctions (f) in FAR3d code is done in terms of sine and cosine components, using real variables:
\begin{eqnarray} 
f(\rho,\theta, \zeta, t) = \sum_{m,n} f^{s}_{mn}(\rho, t) sin(m \theta + n \zeta) \nonumber\\
+ \sum_{m,n} f^{c}_{mn}(\rho, t) cos(m \theta + n \zeta)
\end{eqnarray}
In the following, the cosine component of the eigenfunction is indicated by solid lines and the sine components by dashed lines.

\section{Linear stability analysis \label{sec:lin}}

This section is dedicated to study the linear stability of $n=1$ to $5$ AEs with respect to the EP $\beta$ and energy. The analysis provides information of the EP resonances during the thermalization process, identifying the AE destabilization threshold of different EP populations along the slowing down process, useful information for optimization studies. That way, the model can approach the resonances triggered by a slowing down EP distribution using a set of Maxwellian EP distributions.

\begin{figure}[h!]
\centering
\includegraphics[width=0.45\textwidth]{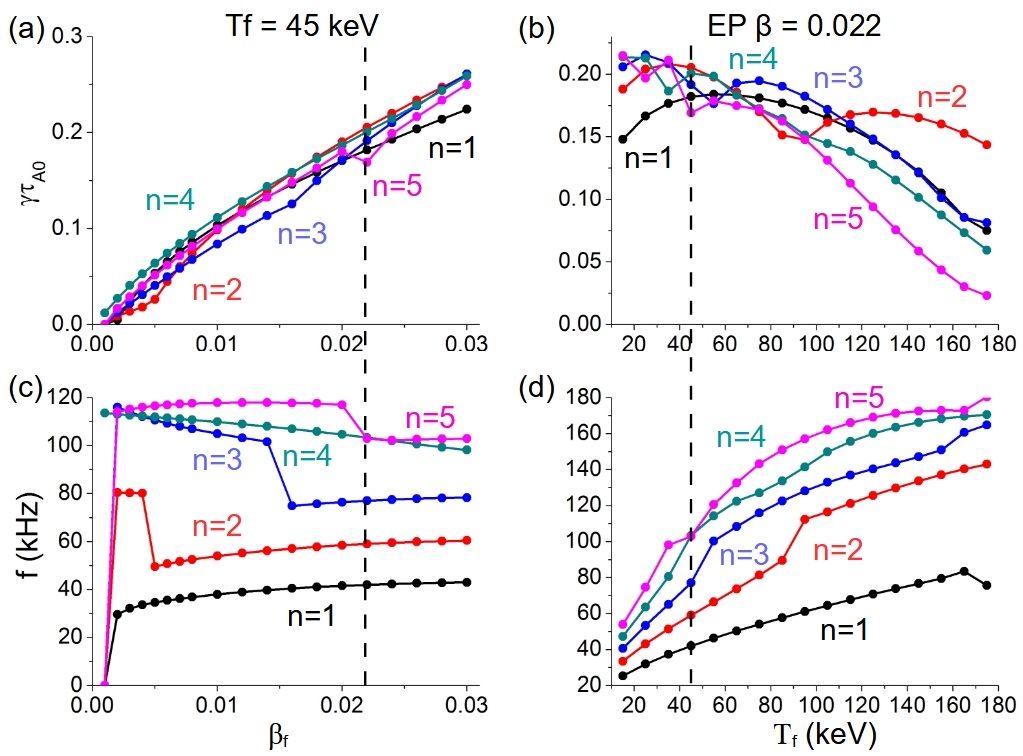}
\caption{Growth rate of $n=1$ to $5$ modes for different values of (a) $\beta_{f}$ and (b) $T_{f}$. Frequency of $n=1$ to $5$ modes for different values of (c) $\beta_{f}$ and (d) $T_{f}$. The dashed vertical black lines indicates the reference case ($\beta_{f} = 0.022$ and $T_{f} = 45$ keV).}\label{FIG:2}
\end{figure}

Figure~\ref{FIG:2} shows the growth rate and frequency of $n=1$ to $5$ modes for different $\beta_{f}$ (fixed $T_{f} = 45$ keV) and $T_{f}$ (fixed $\beta_{f} = 0.022$) values. The destabilization threshold of $n=1$ to $5$ AEs is $\beta_{f} = 0.002$ (panel a and c), except for the $n=4$ AE already unstable for $\beta_{f} = 0.001$. It should be noted that the frequency of $n=1$ to $3$ AEs show a sharp decrease above a given $\beta_{f}$, indicating a transition between dominant modes of different Alfvenic families, particularly from Elliptical AEs (EAEs) to TAEs. Such transition is not observed for the $n=4$ AE. On the other hand, $n=5$ AE shows a transition between EAEs with different dominant poloidal modes, reason why the frequency jump is smaller. The $T_{f}$ analysis indicates the EP energy that leads to the strongest resonance along the slowing down process (largest AE growth rate, panels b and d). For the $n=1$ AE the strongest resonance is observed if $T_{f} = 55$ keV, $35$ keV for the $n=2$ AE, $75$ keV for the $n=3$ AE and $15$ keV for the $n=4$ and $5$ AEs. It should be noted that $n=2$ mode shows a transition between TAE to EAE families for $T_{f} = 95$, $n=3$ for $55$ keV, $n=4$ and $5$ for $45$ keV. The frequency of $n=1$ to $3$ AEs calculated in the simulations is similar to the frequency range of the three instabilities observed in the experiment, although $n=4$ and $5$ AEs have a frequency above the frequency range of the sampled instabilities by the Mirnow coils in the experiment \cite{28}.

\begin{figure*}[h!]
\centering
\includegraphics[width=0.9\textwidth]{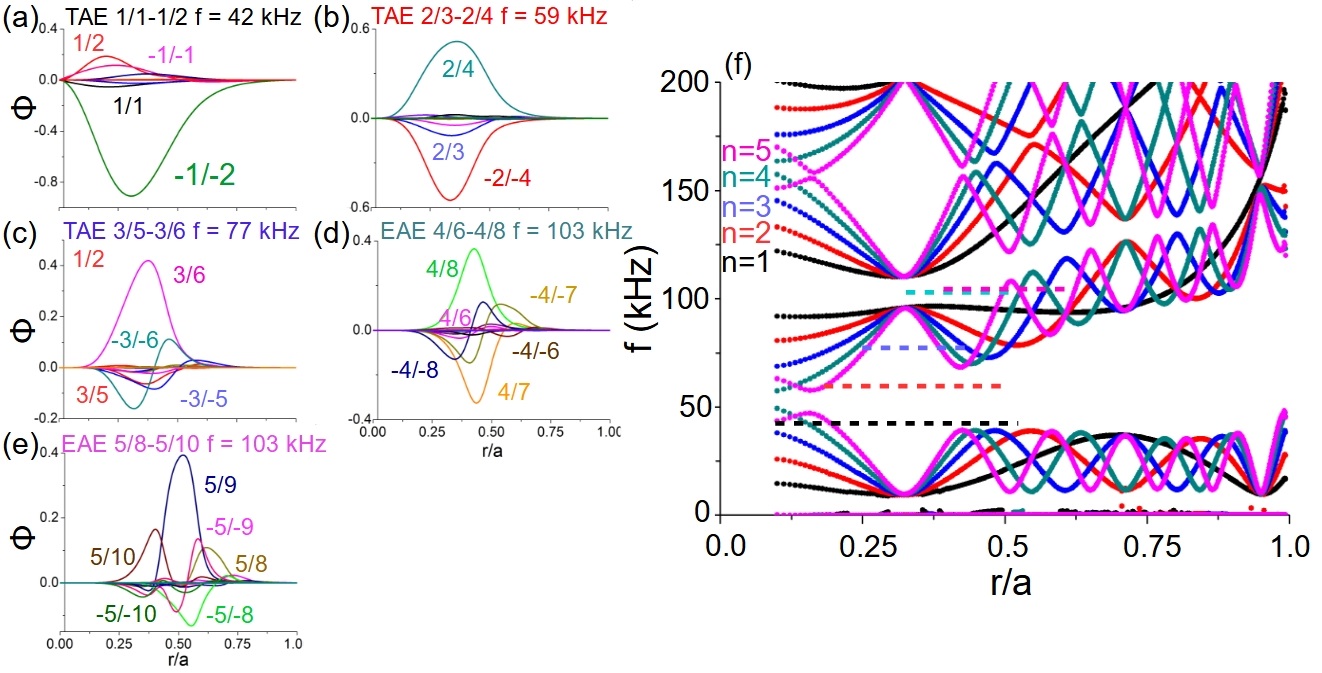}
\caption{Eigenfunction of (a) $1/1-1/2$ TAE, (b) $2/3-2/4$ TAE, (c) $3/5-3/6$ TAE, (d) $4/6-4/8$ EAE and (e) $5/8-5/10$ EAE. (f) Radial location and frequency range of the $n=1$ to $5$ AEs in the continuum gaps for the reference case ($\beta_{f} = 0.022$ and $T_{f} = 45$ keV).}\label{FIG:3}
\end{figure*}

Figure~\ref{FIG:3} shows the eigenfunction and the location of $n=1$ to $5$ AEs in the Alfven gaps for the reference case. $n=1$ to $3$ modes are TAEs destabilized in the inner plasma region, particularly $1/1-1/2$ TAE with $f = 42$ kHz (panel a), $2/3-2/4$ TAE with $f = 59$ kHz (panel b) and $3/5-3/6$ TAE with $f = 77$ kHz (panel c). On the other hand, $n=4$ to $5$ modes are EAEs triggered in the middle plasma region, $4/6-4/8$ EAE with $f=104$ kHz (panel d) and $5/8-5/10$ with $f=103$ kHz (panel e).

\begin{figure}[h!]
\centering
\includegraphics[width=0.45\textwidth]{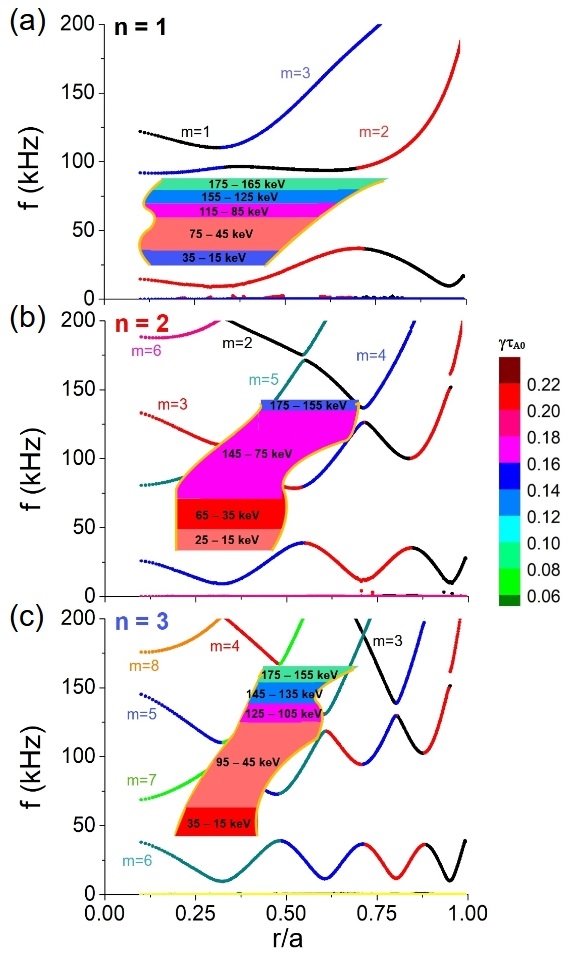}
\caption{Frequency range, radial location and growth rate of $n=1$ (a), $n=2$ (b) and $n=3$ (c) AEs destabilized along the EP slowing down process. Continuum line colors indicate different poloidal modes for each toroidal family. The range of EP energies in the simulations is indicated with respect to the mode growth rate. Simulation EP $\beta$ fixed to $0.022$}\label{FIG:4}
\end{figure}

Figure~\ref{FIG:4} shows the frequency range, radial location and growth rate of the $n=1$ to $3$ AEs destabilized along the EP slowing down process (fixed EP $\beta = 0.022$). The simulations indicate an overlapping of the $n=1$ to $3$ TAEs in the inner plasma region ($r/a = 0.2 - 0.45$) in the frequency range of $40 - 70$ kHz. The largest overlapping is triggered by EPs with energies around $45$ keV (the highest growth rates). The TAEs overlapping must lead to important nonlinear interactions, thus nonlinear simulations are required to study the destabilization of MHD burst. It should be noted that sub-dominant modes are also linearly unstable in the inner plasma region at the frequency range of $40 - 70$ kHz, although the growth rate of these modes is $2$ to $10$ times smaller with respect to the dominant modes. A detail description of these modes is included in the appendix A.

\section{Nonlinear evolution of the TAEs \label{sec:nlin}}

\begin{figure}[h!]
\centering
\includegraphics[width=0.35\textwidth]{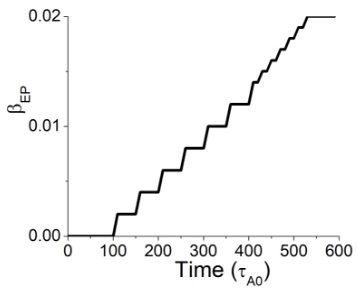}
\caption{EP $\beta$ along the simulation.}\label{FIG:5}
\end{figure}

This section is dedicated to analyze the nonlinear evolution of the TAEs, particularly the destabilization of the MHD burst. On that aim, a nonlinear simulation is performed including EP FLR damping effects (EP Larmor radius is $0.024$ m). The EP $\beta$ increases during the simulation to reproduce the EP population growth along the discharge caused by the tangential NBI, identifying the EP $\beta$ threshold required to destabilize the MHD burst. It should be noted that the increase of the simulation EP $\beta$ is linked to a larger EP density because the EP energy is fixed. Figure~\ref{FIG:5} shows the EP $\beta$ along the simulation. A larger EP $\beta$ in the simulation represents an increase of the EP population (EP density). The EP energy in the simulation is fixed, that is to say, the analysis is focused in the strongest resonance identified by the linear study. It should be noted the EP radial profile evolves during the simulation, thus a change of the EP $\beta$ modifies the scaling of the EP density profile.

\begin{figure}[h!]
\centering
\includegraphics[width=0.45\textwidth]{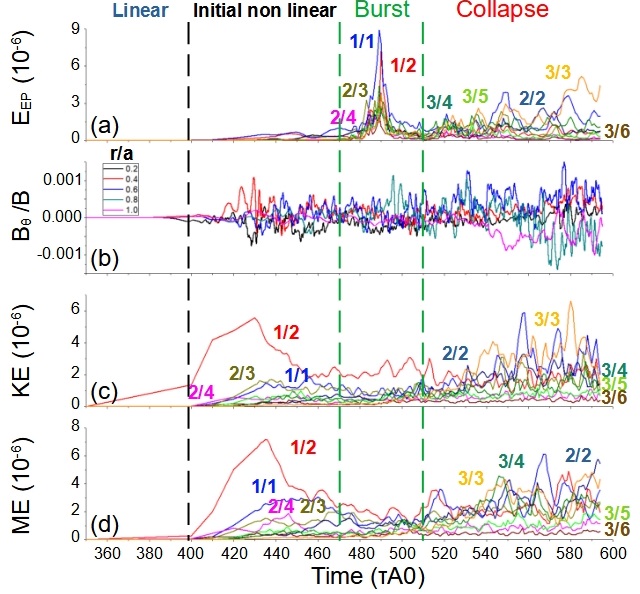}
\caption{Temporal evolution of $n=1$ to $3$ TAE energy (a), poloidal magnetic field perturbation (b), kinetic energy (c) and magnetic energy (d). The line color in the panels a, c and d indicates different mode numbers. The line color in the panel b shows the radial location of the perturbation: black $r/a=0.2$, red $r/a=0.4$, blue $r/a=0.6$, cyan $r/a=0.8$, pink $r/a=1.0$.}\label{FIG:6}
\end{figure}

Figure~\ref{FIG:6} indicates the time evolution of the EP perturbation energy ($E_{EP} = n_{EP} v_{EP}^{2}$ with $n_{EP}$ the EP density and $v_{EP}$ the EP parallel velocity), the perturbation of the poloidal component of the magnetic field ($B_{\theta}$), the kinetic energy ($KE = n (v_{r}^{2} + v_{\theta}^{2})$ with $n$ the thermal plasma density, $v_{r}$ the perturbation of the thermal plasma radial velocity and $v_{th}$ the perturbation of the thermal plasma poloidal velocity) and the magnetic energy ($ME = B_{r}^{2} + B_{\theta}$ with $B_{r}$ the perturbation of the radial component of the magnetic field) in the nonlinear simulation. It should be mentioned that the EP energy is a third order perturbation while KE and ME are second order perturbations, thus the magnitude of EP and thermal plasma energies cannot be directly compared. Consequently, these graphs just provide qualitative information of the perturbation induced in the thermal plasma by the unstable AEs. $1/1-1/2$ TAE is destabilized at $t = 350\tau_{A0}$ (EP $\beta = 0.012$), identified as large $B_{\theta}$ perturbations in the inner-middle plasma region (black and red lines, $r/a = 0.2-0.4$, panel b) and a sudden increase of $1/1$ and $1/2$ modes $KE$ and $ME$ (panels c and d). The graphs also show the simulation overshooting during the transition between the linear and initial nonlinear phases, leading to large peaks of the KE, ME and $B_{\theta}$ perturbations. The simulation recovers from the overshooting at $t = 450\tau_{A0}$. The middle-outer plasma region (blue and pink lines $r/a = 0.6 – 0.8$) is destabilized during the initial nonlinear phase (lasting until $t = 460\tau_{A0}$) pointing out the outward propagation of $1/1-1/2$ TAE. The local maxima of $1/1$ and $1/2$ modes $KE$ and $ME$ is observed around $t = 430\tau_{A0}$, indicating an energy transfer from $1/1-1/2$ TAE to the thermal plasma that induces the generation of zonal flows and currents. The bursting phase begins at $t = 470\tau_{A0}$ once EP $\beta = 0.017$. There is a sudden increase of $E_{EP}$ linked to the $1/1$ mode (panel a), followed by a large peak of $n=1$ to $3$ modes $E_{EP}$ at $t = 490\tau_{A0}$ identified as the MHD burst. At the beginning of the bursting phase the middle plasma region is strongly destabilized although, after the MHD burst, the perturbation moves towards the periphery. In the experiment the TAEs are stable after the MHD burst due to a partial loss of the EP population. In the nonlinear simulation, TAEs are marginal stable after the MHD burst if the EP $\beta$ is smaller than $0.01$. Present simulation shows an hypothetical scenario obtained by further increasing the EP $\beta$ to $0.02$ after the MHD burst, that is to say, if the EP perturbation is reinforced. This hypothetical phase is called Collapse phase and it is dedicated to study the plasma stability if the MHD burst does not causes a partial loss of the EP population. During the collapse phase the perturbation is mainly located between the middle-outer plasma region and the dominant modes resonate at the plasma periphery, particularly $1/1$ mode and overtones showing successive local maxima of $E_{EP}$, $ME$ and $KE$.

\begin{figure}[h!]
\centering
\includegraphics[width=0.45\textwidth]{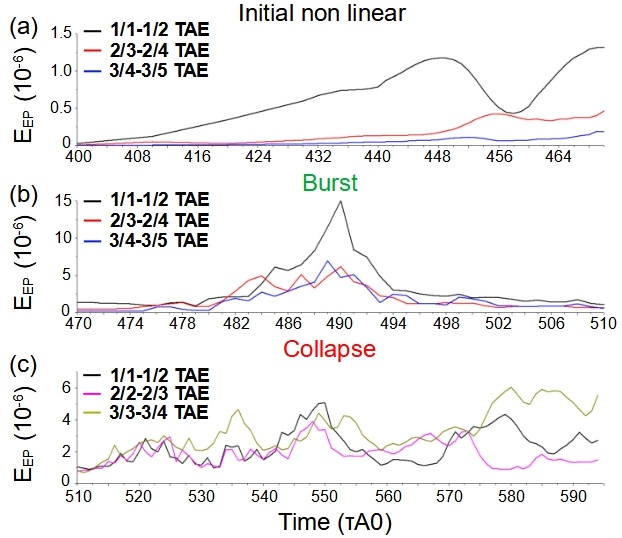}
\caption{Temporal evolution of $n=1$ to $3$ TAEs energy during the (a) initial nonlinear, (b) bursting and (c) collapse phases. The line color indicates different TAEs: black $1/1-1/2$, red $2/3-2/4$, blue $3/4-3/5$, pink $2/2-2/3$ and dark yellow $3/3-3/4$.}\label{FIG:7}
\end{figure}

Figure~\ref{FIG:7} shows the $E_{EP}$ evolution of $n=1$ to $3$ TAEs during the initial nonlinear, bursting and collapse phases. During the initial nonlinear phase $1/1-1/2$ TAE is first destabilized followed by $2/3-2/4$ TAE (panel a). There is a nonlinear energy transfer from $1/1-1/2$ TAE to $2/3-2/4$ TAE, particularly large between $t = 450 – 460 \tau_{A0}$, leading to the further destabilization of $2/3-2/4$ TAE. At the beginning of the bursting phase $1/1-1/2$ TAE is the most energetic mode (panel b), although the nonlinear energy channeling towards $2/3-2/4$ and $3/4-3/5$ TAEs leads to a dominant $2/3-2/4$ TAE at $t = 482 \tau_{A0}$. From $t = 485 \tau_{A0}$ the $1/1-1/2$ TAE is again the dominant perturbation showing an energy $6$ times larger compared to the energy at the beginning of the bursting phase. The local maxima of $n=1$ to $3$ TAEs energy is observed around $t = 490 \tau_{A0}$ once the MHD burst is triggered. After the MHD burst the energy of $n=1$ to $3$ TAEs decreases, that is to say, $n=1$ to $3$ TAEs are weakened. Regarding the collapse phase (panel c), the dominant poloidal modes of $n=2$ and $n=3$ TAEs evolve to $2/2-2/3$ and $3/3-3/4$. There are several local maxima of the TAEs $E_{EP}$, particularly at $t = 535 \tau_{A0}$ by $3/3-3/4$ TAE, at $t = 550 \tau_{A0}$ by $1/1-1/2$ TAE and at $t = 567 \tau_{A0}$ by $2/2-2/3$ TAE. From $t = 575 \tau_{A0}$ the $3/3-3/4$ TAE shows the largest $E_{EP}$ until the simulation ends due to the destabilization of pressure gradient driven modes at TAEs with large growth rates at the plasma periphery.

\begin{figure}[h!]
\centering
\includegraphics[width=0.45\textwidth]{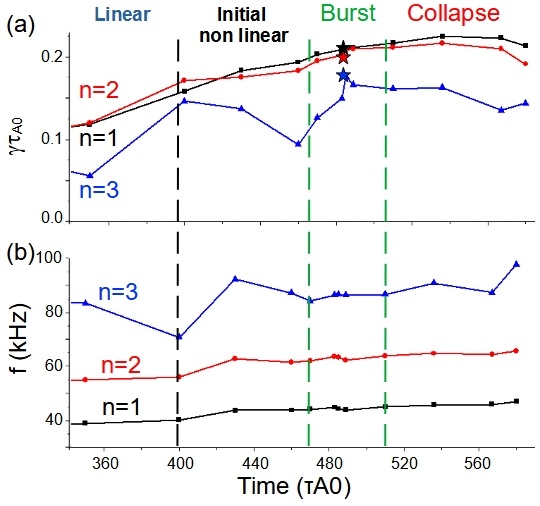}
\caption{(a) Growth rate and (b) frequency of the dominant $n=1$ to $3$ AEs calculated by linear simulations using the equilibria evolved along the nonlinear simulation.}\label{FIG:8}
\end{figure}

Figure~\ref{FIG:8} indicates the growth rate and frequency of $n=1$ to $3$ TAEs calculated by linear simulations using the equilibria evolved along the nonlinear simulation. The equilibrium profiles evolve along the nonlinear simulation, for example the plasma pressure, EP density and iota radial profiles, modifying too the Alfven continuum gaps and the continuum damping effects. Consequently, TAE stability changes. This analysis provides an approximate evaluation of the stabilizing or destabilizing effects associated with the nonlinear interaction between AEs of different toroidal families, as well as between AEs and the thermal plasma. The growth rate of $n=1$ to $3$ AEs calculated for different EP $\beta$ values fixed the EP energy to $45$ keV (fig~\ref{FIG:2}a) is compared with the growth rate of linear simulations using the profiles of the nonlinear simulations (for simulations with the same EP $\beta$). The growth rate of $n=1-3$ TAEs is similar in the linear phase of the nonlinear simulation with respect to the linear simulation performed in the previous section (considering the EP $\beta$ increment along the nonlinear simulation), showing an slightly increase around $10 - 15 \%$ for $n=1$ and $2$ TAEs. Along the initial nonlinear phase the growth rate is $30 \%$ larger for $n=1$ to $2$ TAEs and $10 \%$ larger for $n=3$ TAEs compared to the linear simulations for the same EP $\beta$, pointing out a further destabilization of the modes caused by nonlinear energy transfers. There is a further enhancement of the growth rate in the bursting phase, increasing to $35 \%$ for $n=1$ to $2$ TAEs and $25 \%$ for $n=3$ TAE, unveiling the key role of the nonlinear interaction between modes during the MHD burst. On the other hand, the growth rate only increases by $15 \%$ during the collapse phase for $n=1$ to $2$ TAEs, that is to say, nonlinear effects on the Collapse phase have a smaller role. It should be noted that $n=3$ TAE growth rate decreases during the initial nonlinear and Collapse phases, explained by the effect of the continuum gaps evolution on the mode stability. $n=3$ TAE is destabilized nearby the upper bound of the TAE gap, thus $n=3$ TAE growth rate may decrease if the profiles evolution leads to a slender TAE gap, enhancing the continuum damping and reducing the $n=3$ TAE radial width.

In summary, the nonlinear simulation during the initial nonlinear and bursting phases reproduces the frequency range of the instabilities measured in the experiment, consistent with the destabilization of $n=1$ to $3$ TAEs. In the following, the analysis of the initial nonlinear and bursting phases is performed independently with respect to the Collapse phase, separating the study dedicated to reproduce the experimental observations and the hypothetical scenario represented by the Collapse phase.

\subsection{Experiment simulation: MHD burst}

\begin{figure}[h!]
\centering
\includegraphics[width=0.45\textwidth]{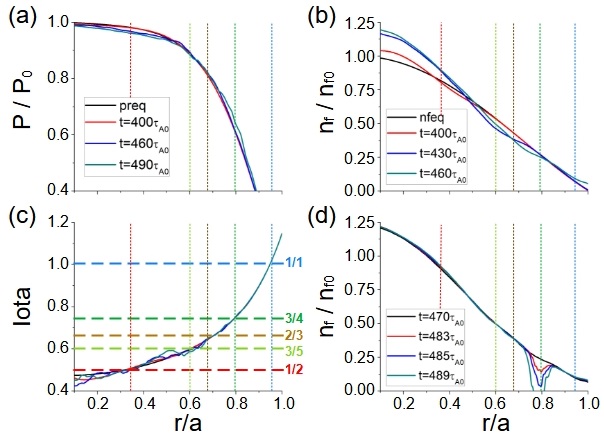}
\caption{Temporal evolution of the (a) normalized pressure, (b) normalized EP density during the initial nonlinear phase, (c) iota and (d) normalized EP density during the bursting phase. The vertical dashed lines indicates the radial location of the main resonant rational surfaces: red $1/2$, light green $3/5$, dark yellow $2/3$, green $3/4$ and blue $1/1$.}\label{FIG:9}
\end{figure}

Figure~\ref{FIG:9} shows the evolution of the iota, EP density and pressure profiles during the initial nonlinear and bursting phases. The pressure profile in the initial nonlinear phase (panel a) indicates small deviations away from the equilibrium caused by the resonant $1/2$ rational surface in the inner plasma region. Likewise, the deviations of the iota profile (panel c) are small, mainly located in the middle plasma region and nearby the magnetic axis, generated by local electric fields induced by the TAEs. On the other hand, the EP density profile shows important deviations with respect to the the equilibrium; EP density increases in the inner plasma region (panel b) and shows flattenings induced by $1/1-1/2$ TAE around $r/a = 0.35$ ($t = 400 \tau_{A0}$) and by $2/3-2/4$ TAE ($t = 430 \tau_{A0}$) and $3/4-3/5$ TAE ($t = 460 \tau_{A0}$) in the middle-outer plasma region. Regarding the bursting phase, once the MHD burst is triggered (panel d), the EP density profile shows a large decrease around $r/a = 0.75 – 0.8$, plasma region where the $3/4$ rational surface is resonant, leading to a radial redistribution of the EP density that can indicate a partial lost of the EP population.

\begin{figure}[h!]
\centering
\includegraphics[width=0.45\textwidth]{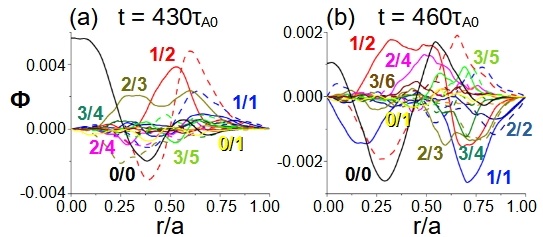}
\caption{Electrostatic potential perturbation at (a) $t = 430 \tau_{A0}$ and (b) $t = 460 \tau_{A0}$.}\label{FIG:10}
\end{figure}

Figure~\ref{FIG:10} shows the electrostatic potential eigenfunction during the initial nonlinear phase. At $t = 430 \tau_{A0}$ (panel a) $1/1-1/2$ TAE is unstable in the middle plasma region. $2/3-2/4$ TAE is also destabilized, partially overlapped with $1/1-1/2$ TAE, leading to a nonlinear energy transfer from $n=1$ to $n=2$ TAE. It should be noted the $1/1-1/2$ TAE amplitude is particularly large in the middle plasma region, two times larger compared to the $2/3-2/4$ TAE, consistent with Figs 7a and 8a that indicate $1/1-1/2$ TAE is the dominant instability. The large amplitude of $1/1-1/2$ and $2/3-2/4$ TAEs is linked to the simulation overshooting observed in the transition between the linear and nonlinear phases, also showing a large overlapping between the modes. In addition, the eigenfunction of mode $0/0$ shows its maximum amplitude near the magnetic axis, as well as secondary local maxima overlapped with the eigenfunction amplitude maxima of $n=1$ and $2$ TAEs, pointing out an energy transfer from the TAEs to the thermal plasma. Again, this is consistent with the large perturbation of the thermal plasma observed in the figs. 6c and 6d, leading to a local maxima of the KE and ME. At $t = 460 \tau_{A0}$ (panel b) $3/4-3/5$ TAE is also unstable at the middle-outer plasma region, partially overlapped with $1/1-1/2$ TAE that now covers main part of the normalized minor radius, thus there is a nonlinear energy channeling from $n=1$ to $n=3$ TAE. In addition, $2/3-2/4$ TAE displaces outwards, now located between the middle-outer plasma region. The $1/1-1/2$ TAE amplitude is smaller compared to $t = 430 \tau_{A0}$ although $2/3-2/4$ TAE amplitude increases, pointing out the overlapping and energy transfer between $n=1$ and $n=2$ TAEs is smaller because $n=1$ TAE energy is lower (see fig. 7a).

\begin{figure}[h!]
\centering
\includegraphics[width=0.45\textwidth]{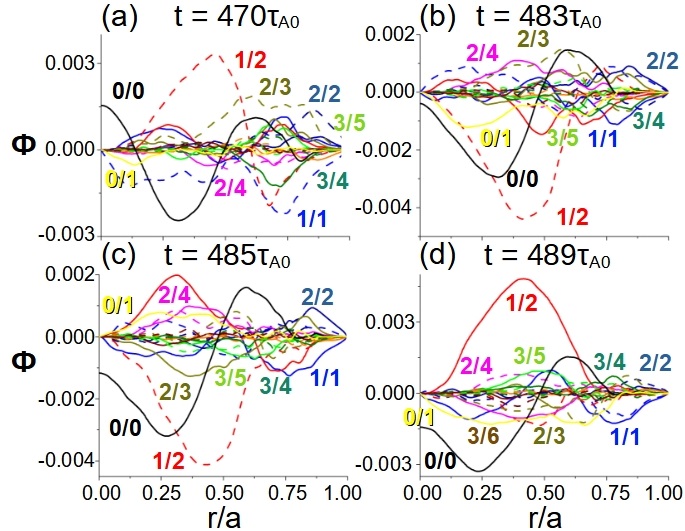}
\caption{Electrostatic potential perturbation at (a) $t = 470 \tau_{A0}$, (b) $t = 483 \tau_{A0}$, (c) $t = 485 \tau_{A0}$ and (d) $t = 489 \tau_{A0}$.}\label{FIG:11}
\end{figure}

Figure~\ref{FIG:11} shows the electrostatic potential eigenfunction during the bursting phase. At $t = 470 \tau_{A0}$ (panel a), $1/1-1/2$ TAE covers main part of the plasma, overlapped with $2/3-2/4$ and $3/4-3/5$ TAEs located in the middle-outer plasma region, thus the nonlinear energy channeling from $n=1$ to $n=2$ and $3$ TAEs continues. The TAEs amplitude increases compared to $t = 460 \tau_{A0}$ (fig 10b), indicating an enhancement of the overlapping between the modes. Once the MHD burst is triggered, $t = 483 \tau_{A0}$ (panel b), the local maxima of $2/3-2/4$ and $1/1-1/2$ eigenfunctions amplitude overlap in the middle plasma region, leading to a large energy transfer from $n=1$ to $n=2$ TAEs. Consequently, $2/3-2/4$ TAE is further destabilized evolving to the dominant perturbation (see fig.~\ref{FIG:7}b, red line). In addition, part of $n=1$ TAE energy is channeled towards the thermal plasma leading to the further destabilization of the modes $0/0$ and $0/1$ showing the maximum amplitude of the eigenfunction in the inner plasma region. The increment of $1/1-1/2$ TAE amplitude indicates an enhancement of the TAE overlapping. At $t = 485 \tau_{A0}$ (panel c), the local maxima of $3/4-3/5$ and $1/1-1/2$ eigenfunctions amplitude overlap, enhancing the energy transfer from $n=1$ to $n=3$ TAE a leading to the further destabilization of $3/4-3/5$ TAE (see fig.~\ref{FIG:7}b, blue line). The amplitude of $1/1-1/2$ TAE remains in maximum values showing a strong mode overlapping. At $t = 489 \tau_{A0}$ (panel d) the eigenfunction amplitude maxima of $1/1-1/2$, $2/3-2/4$ and $3/4-3/5$ TAEs are aligned, leading to a large destabilizing feedback between TAEs, consistent with the local maxima of $E_{EP}$ once the MHD burst is triggered. The amplitude of all TAEs is maxima, particular for the $1/1-1/2$, pointing out the reinforcement of the mode overlapping. Also, $0/0$ and $0/1$ modes overlap with $1/1-1/2$ TAE, thus the energy transfer towards the thermal plasma continues. Consequently, the simulation indicates the MHD burst is caused by the overlapping between $n=1$ to $3$ TAEs.

\begin{figure}[h!]
\centering
\includegraphics[width=0.45\textwidth]{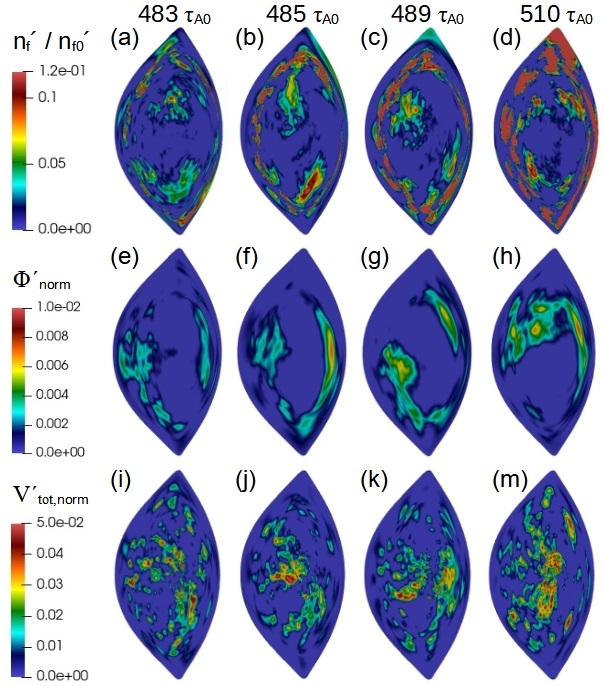}
\caption{Poloidal contour of the normalized perturbations of the EP density (a to d), electrostatic potential (e to h) and thermal plasma velocity (i to m) during the MHD burst.}\label{FIG:12}
\end{figure}

Figure~\ref{FIG:12} shows the evolution of the poloidal contour of the normalized EP density, electrostatic potential and thermal plasma velocity perturbations during the MHD burst. The poloidal contour of the normalized EP density shows a complex perturbation pattern between the inner and outer plasma region at $t = 483 \tau_{A0}$ induced by $n=1$ to $n=3$ TAEs, although two perturbations generated by $1/1-1/2$ TAE are dominant in the middle plasma region (panel a). The mode overlapping that triggers the MHD burst is identified as the perturbation that connects the inner and outer plasma regions at $t = 485 \tau_{A0}$ (panel b). The EP density perturbation driven by $n=1$ to $3$ TAEs at $t = 489 - 510 \tau_{A0}$ indicates a redistribution and partial lost of the EP population (panels c and d). The evolution of the electrostatic potential perturbation shows the generation of zonal currents (panels e to h) leading to local modifications of the iota profile in the middle plasma region and near the magnetic axis (see graph 8c). Shear flows are also generated, particularly between the inner and middle plasma region (panels i to m). The perturbation of the radial component of the thermal plasma velocity extends from the inner to the outer plasma region, although the perturbation of the poloidal component is mainly located in the inner plasma region.

In summary, the initial nonlinear and bursting phases in the simulation reproduce several features of the MHD bursts observed in the LHD experiment, for example the frequency range of the instabilities measured along the discharge and the EP population losses induced by the AEs. The MHD burst is triggered once the EP energy reaches a local maxima (strong $n=1$ to $3$ TAEs destabilization), a local maxima of the KE and ME (large energy transfer to the thermal plasma) and $B_{\theta}$ oscillations around $r/a = 0.6-0.8$ (strong perturbation of the plasma in the middle-outer plasma region). In addition, there is a maxima of the $n=1-3$ TAEs amplitude peaking at the same radial location. Consequently, the analysis concludes the MHD burst is caused by the overlapping between $n=1$ to $3$ TAEs. The overlapping between TAEs is promoted by the nonlinear energy transfer from $n=1$ TAE towards $n=2$ and $3$ TAEs, leading to the modes further destabilization. The MHD burst causes a radial redistribution of the EP density, particular in the middle-outer plasma region, that can be explained as a partial lost of the EP population. Another consequence of the MHD burst is the generation of shear flows and zonal currents by the TAEs, disturbing the thermal plasma flows as well as the iota profile due the generation of local and time evolving electric fields.

\subsection{Hypothetical scenario: collapse of the plasma periphery}

The simulation indicates the stabilization of the TAE after the MHD burst if the EP $\beta$ is below the TAE destabilization threshold, EP $\beta$ decrease that is consistent with the loss of a fraction of the EP population during the MHD burst (see panel a and e of figure 9 in the reference \cite{28}). In this section, we consider an hypothetical scenario characterized by a smaller fraction of EP losses after the MHD burst combined with an enhancement of the NBI injection, leading to a further enhancement of the EP $\beta$, not a decrease. Such scenario is not observed in LHD plasma although it provides useful information for future devices with a stronger EP destabilization effect and improved EP confinement compared to LHD.

\begin{figure}[h!]
\centering
\includegraphics[width=0.3\textwidth]{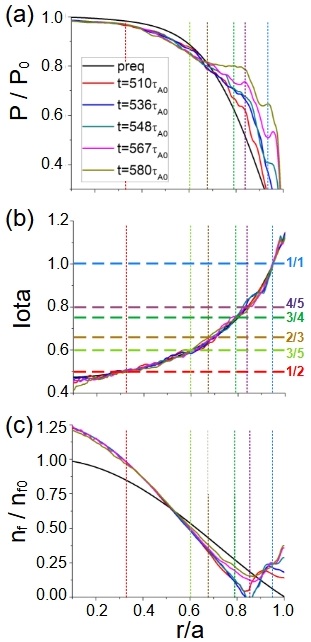}
\caption{Temporal evolution of the (a) normalized pressure, (b) iota and (c) normalized EP density during the collapse phase. The vertical dashed lines indicates the radial location of the main resonant rational surfaces: red $1/2$, light green $3/5$, dark yellow $2/3$, green $3/4$ and blue $1/1$.}\label{FIG:13}
\end{figure}

Figure~\ref{FIG:13} shows the evolution of iota, EP density and pressure profiles in the Collapse phase. A large pressure gradient builds up at the plasma periphery (panel a) due to wide pressure flattenings induced at $r/a = 0.7-0.85$ nearby the resonant rational surfaces $2/3$, $3/4$ and $4/5$, as well as near the $1/1$ around $r/a = 0.9-0.95$. The main perturbations of the iota profile are located between the middle-outer plasma region (panel b). The EP density increases at the plasma periphery, thus the EP population recovers after the burst (panel c). In addition, there is an inversion of the EP density profile at $r/a > 0.85$ caused by the outward flux of EPs. 

\begin{figure}[h!]
\centering
\includegraphics[width=0.45\textwidth]{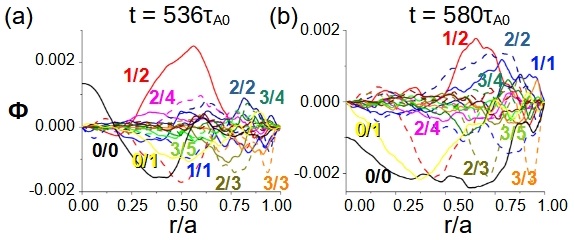}
\caption{Electrostatic potential perturbation at (a) $t = 536 \tau_{A0}$ and (b) $t = 580 \tau_{A0}$.}\label{FIG:14}
\end{figure}

Figure~\ref{FIG:14} shows the electrostatic potential eigenfunction during the Collapse phase. There is a transition from dominant $2/3-2/4$ and $3/4-3/5$ TAEs in the bursting phase to dominant $2/2-2/3$ and $3/3-3/4$ TAEs, that is to say, the plasma periphery is further destabilized (panel a). $1/1-1/2$ TAE covers main part of the plasma and continues feeding energy to $n=2$ and $3$ TAEs, moving towards the periphery as the Collapse phase evolves (panel b). In addition, the energy transfer from TAEs to the thermal plasma enhances, thus the amplitude of $0/0$ and $0/1$ modes increases, also extended towards the periphery as the Collapse phase advances. It should be noted the mode amplitude is smaller compared to the busting phase, thus the overlapping between TAEs is weaker.

\begin{figure}[h!]
\centering
\includegraphics[width=0.3 \textwidth]{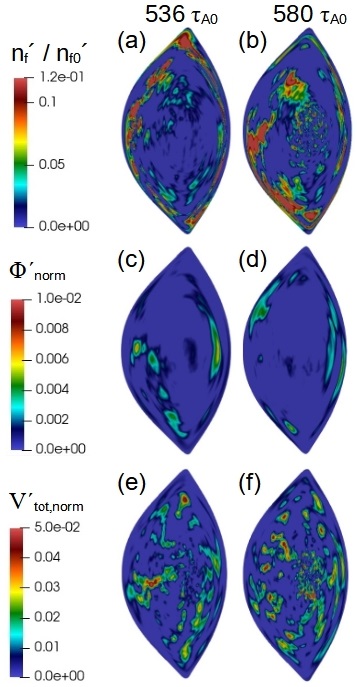}
\caption{Poloidal contour of the normalized EP density (a and b), electrostatic potential (c and d) and thermal plasma velocity (e and f) perturbations during the collapse phase.}\label{FIG:15}
\end{figure}

Figure~\ref{FIG:15} shows the evolution of the poloidal contour of the normalized EP density, electrostatic potential and thermal plasma velocity perturbations during the Collapse phase. The EP density perturbation is located at the plasma periphery, further enhanced along the Collapse phase (panels a and b). The electrostatic field perturbation is smaller with respect to the bursting phase, mainly located at the plasma periphery (panels c and d). Zonal flows are also generated (panels e and f).

\begin{figure}[h!]
\centering
\includegraphics[width=0.45\textwidth]{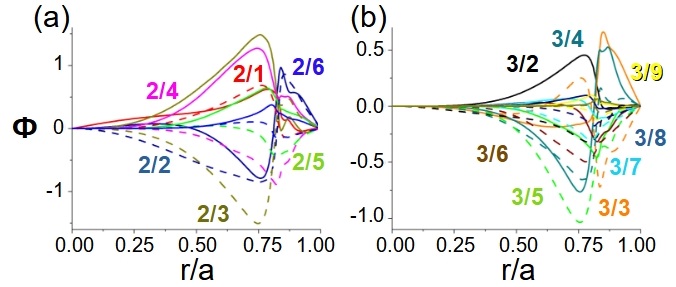}
\caption{Electrostatic potential perturbation of linear simulations using the profiles of the nonlinear simulation at $t = 536 \tau_{A0}$ for (a) $n=2$ and (b) $n=3$ modes.}\label{FIG:16}
\end{figure}

The simulation terminates after the destabilization of low $n$ ballooning modes and TAEs with large growth rates at the plasma periphery. Figure~\ref{FIG:16} show the eigenfunction of the $n=2$ and $3$ dominant modes for linear simulations performed using the profiles of the nonlinear simulation at $t = 548 \tau_{A0}$. The perturbations are ballooning modes triggered at the plasma periphery.

Consequently, the hypothetical Collapse phase highlights the interconnection between EP density and pressure gradient driven modes, leading to the destabilization and collapse of the plasma periphery.

\section{Conclusions and discussion \label{sec:conclusions}}

Linear and nonlinear simulations are performed using FAR3d code, dedicated to analyze the stability of TAEs in LHD discharges, particularly the triggering of the MHD burst. 

The linear simulations indicate the destabilization of $1/2-1/1$, $2/3-2/4$ and $3/5-3/6$ TAEs between the inner-middle plasma region in the frequency range of $45-75$ kHz, consistent with the experiment observations. $n=1$ to $3$ TAEs are triggered by the energetic particle population with an energy of $45$ keV and EP $\beta = 0.022$, consistent with MORH and MEGA EP models \cite{61}.

\begin{figure*}[h!]
\centering
\includegraphics[width=0.95 \textwidth]{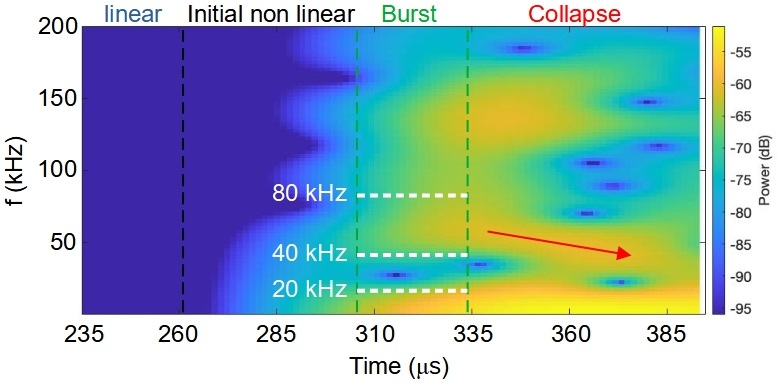}
\caption{Spectrogram of the poloidal component of the magnetic field perturbation along the nonlinear simulation.}\label{FIG:17}
\end{figure*}

The linear stability analysis for different EP energies indicates that there is an overlapping between $n=1$ to $3$ AEs along the slowing down process. In addition, the strongest resonance (AEs with the largest growth rates) is caused by particles with energies around $45$ keV. The EP $\beta$ threshold to destabilize $n=1$ to $5$ AEs is $0.002$, except for $n=4$ AE already unstable for $0.001$. 

The nonlinear simulations show the energy transfer from $1/1-1/2$ TAE towards $2/3-2/4$ and $3/4-3/5$ TAEs, leading to the further destabilization of these modes. In addition, a fraction of $1/1-1/2$ TAE energy is transferred towards the thermal plasma, destabilizing $0/0$ and $0/1$ modes. 

Figure~\ref{FIG:17} shows the spectrogram of the poloidal component of the magnetic field perturbation along the nonlinear simulation. The model reproduces the frequency range of the TAE activity observed in the experiment during the MHD burst, between $40 – 80$ kHz, as well as the pressure gradient driven mode activity below $20$ kHz. As it was already observed in the linear analysis, fig~\ref{FIG:8}, the nonlinear simulation indicates a frequency relation between $n=1$ to $3$ TAEs with $f_{n=2} \approx 1.5 f_{n=1}$ and $f_{n=3} \approx 2 f_{n=1}$. It should be noted that, in general, the linearly unstable AEs do not have to follow the rule $f_{n=2} = 2f_{n=1}$ or $f_{n=3} = 3f_{n=1}$. The frequencies follow the shear Alfven wave dispersion relation $\omega = k_{||} v_{A0}$ with $k_{||} \sim (n – m/q)/R_{0}$. As the toroidal mode number increases the dominant poloidal mode number also increases leading to roughly constant $k_{||}$ and $\omega$ values. This is observed in the Alfven continuum calculations (figure 3f) where the continua of $n=2,3,…$ overlay $n=1$ continua. The continuum boundaries surrounding the open gaps in Fig. 3f introduce strong damping for waves whose frequency and radial location leads to a continuum crossing. This effect channels the frequencies of even non linearly generated Alfven waves into the open frequency gap regions. Since these open frequency regions are similar for all toroidal mode numbers, they limit the nonlinear coupling of mode energy to higher frequencies, thus the same discussion can be done for nonlinear Alfven wave interactions. If $n=2$ AE is destabilized by the nonlinear drive of $n=1$ AE, the frequency of the $n=2$ AE will likely be influenced by the frequency range of the $n=1$ AE, but not necessary must follow the rule $f_{n=2} = 2f_{n=1}$. The spectrogram also shows that, during the MHD burst, the instability is originated in the frequency range of the $50-55$ kHz, later extending to the frequency range between $45$ to $80$ kHz, consistent with the experimental observations. In addition, the model predicts the destabilization of EAEs with frequencies around $125$ kHz in the middle plasma region at the end of the bursting phase.

The EP density profile in the nonlinear simulation phase shows several flattenings induced in the radial location where $n=1$ to $3$ TAEs are destabilized, indicating the instability saturation. A further destabilization of the TAEs due to the increment of the simulation EP $\beta$ triggers the MHD burst, leading to a large drop of the EP density profile around $r/a = 0.75 – 0.8$, radial location where the local amplitude maxima of $n=1$ to $3$ TAEs overlaps, pointing out a radial redistribution of the EP density that can be interpreted as a partial lost of the EP population during the burst. Consequently, the burst is caused by the overlapping between $n=1$ to $3$ TAEs.

The nonlinear simulation shows the generation of shear flows and zonal current nearby the TAEs. The perturbation of the radial and poloidal components of the thermal plasma velocity increases during the burst. Unstable TAEs can enhance the radial fluxes of the thermal plasma at the middle-outer plasma region and generate shear flows in the inner-middle plasma region. The zonal currents are relatively weaker, leading to small deviations from the equilibrium iota profile.

The hypothetical Collapse phase is also analyzed, representing an scenario where the MHD burst does not lead to a large reduction of the EP population and TAE stabilization, that is to say, the EP $\beta$ of the simulations is fixed and remains above the EP $\beta$ threshold required to destabilize the TAEs. The analysis of this hypothetical scenario shows the destabilization of $2/2-2/3$ and $3/3-3/4$ TAEs at the plasma periphery, as well as the generation of a pedestal due to the built up of large pressure gradient that overcomes the stability limit of ballooning modes. The strong destabilization of the plasma periphery leads to the plasma collapse, showing some common features with the trigger mechanism of the core density collapses also observed in LHD plasma for outward shifted configurations \cite{62,63}.

It should be noted that the present study does not provide information of the EP transport during the burst MHD events, a topic that will be discussed in future analysis. Nevertheless, the EP fluxes induced by the nonlinear coupling of AEs destabilized by multiple toroidal families may lead to an avalanche-like EP transport, that is to say, the plasma may show characteristics of a self-organized criticality system \cite{64,65}. EP avalanches is one of the concerns regarding future nuclear fusion reactors, potentially leading to large losses of alpha particles before thermalization, increasing the operational requirements and hampering the economical viability of the reactor.  

Present simulations show the hazardous consequences of AE overlapping on plasma heating efficiency. The AEs radial and frequency range overlapping can trigger bursting events, leading to large EP population losses before thermalization. Nonlinear energy transfers between toroidal families promote the destabilization of AEs and their overlapping, particularly for AEs triggered inside wide continuum gaps with a broad eigenfunction width. Different strategies exist to avoid bursting activity induced by AE overlapping. For example, the minimization of the nonlinear energy transfer if the NBI injection intensity during the discharge is controlled (lower EP $\beta$), that is to say, avoiding the triggering of a strongly unstable AE that can feed energy and destabilize AEs of different toroidal mode families. Another possibility is modifying the magnetic configuration to increase the magnetic shear or the plasma fueling to generated peaked thermal ion density profiles, leading to slender and radially localized continuum gaps that avoids radially extended AEs and their overlapping.

\ack

The authors would like to thank the LHD technical staff for their contributions in the operation and maintenance of LHD. This work was supported by the Comunidad de Madrid under the project 2019-T1/AMB-13648, Comunidad de Madrid - multiannual agreement with UC3M (“Excelencia para el Profesorado Universitario” - EPUC3M14 ) - Fifth regional research plan 2016-2020 and NIFS07KLPH004 

\section*{Appendix A: sub-dominant modes}

\begin{figure}[h!]
\centering
\includegraphics[width=0.45 \textwidth]{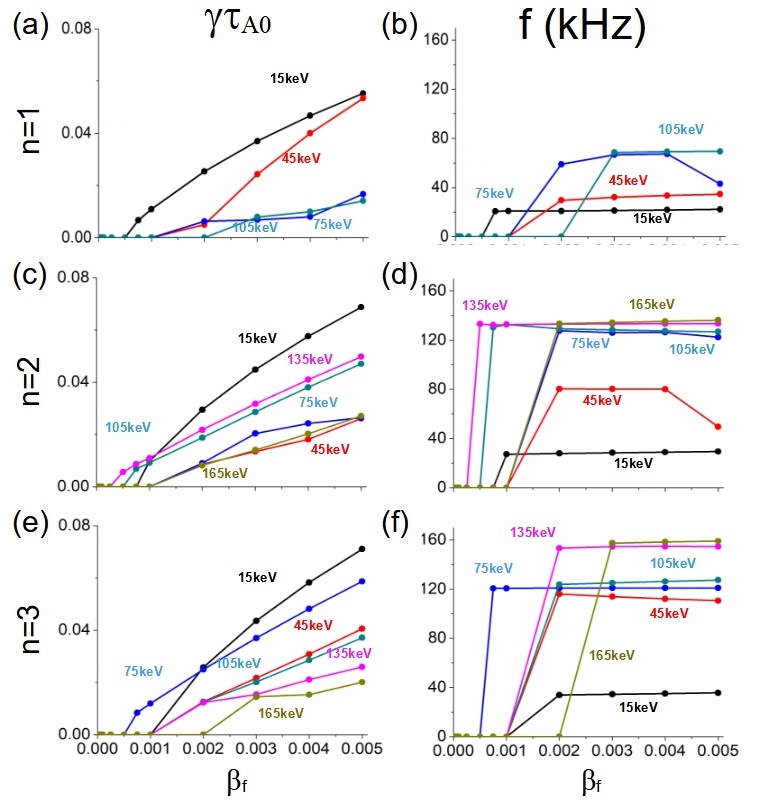}
\caption{Growth rate and frequency of $n=1$ (a and b), $n=2$ (c and d) and $n=3$ (e and f) AEs.}\label{A1}
\end{figure}

\begin{figure}[h!]
\centering
\includegraphics[width=0.45 \textwidth]{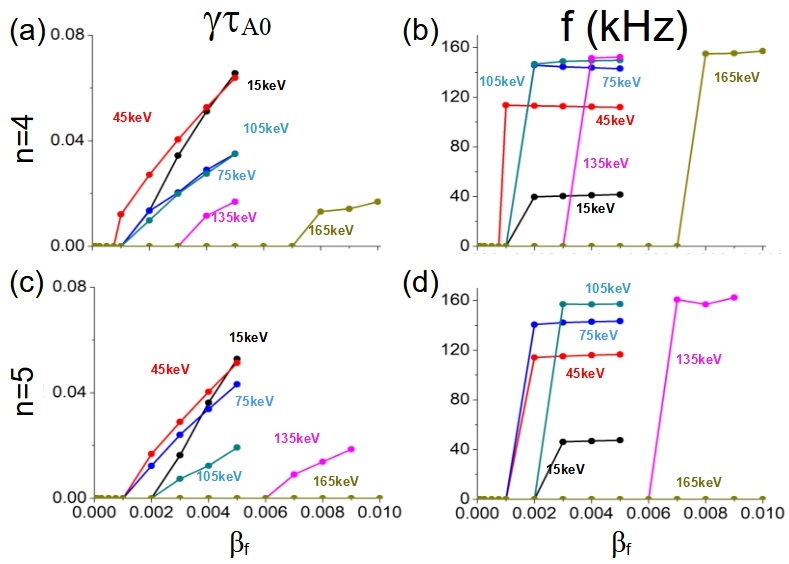}
\caption{Growth rate and frequency of $n=4$ (a and b) and $n=5$ (c and d) AEs.}\label{A2}
\end{figure}

\begin{figure}[h!]
\centering
\includegraphics[width=0.45 \textwidth]{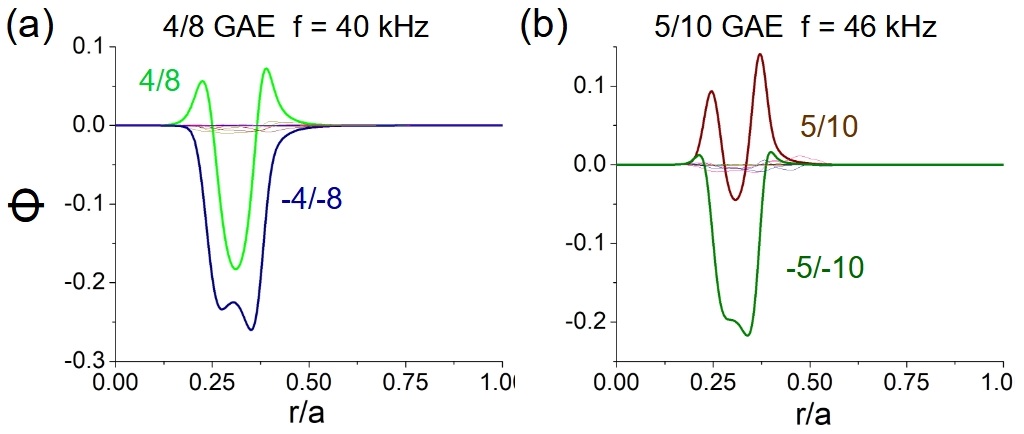}
\caption{Eigenfunction of (a) $n=4$ and (b) $n=5$ sub-dominant AEs for simulations with EP $\beta = 0.005$ and $T_{f} = 15$ keV.}\label{A3}
\end{figure}

This appendix is dedicated to analyze the sub-dominant modes identified in the linear simulations. Figure~\ref{A1} and~\ref{A2} show the growth rate and frequency of $n=1$ to $5$ AEs for different EP $\beta$ and $T_{f}$ values, showing growth rates $2$ to $10$ times smaller compared to the dominant modes. Low energy EPs ($T_{f} = 15$ keV) destabilize the sub-dominant AEs with the largest growth rates, $n=1$ to $3$ AEs with frequencies below $40$ kHz. High energy EPs trigger AEs with lower growth rates compared to low energy EP and frequencies above $120$ kHz. It shoould be noted that the growth rate of the sub-dominant AEs triggered is significantly smaller compared to the dominant modes. Also, the frequency range of these sub-dominant modes is different regarding the frequency range of the MHD burst. Thus, the effect of these modes can be neglected in the nonlinear study in first approximation because the overlapping with the dominant modes is weak. Nevertheless, the frequency of $n=4$ and $5$ sub-dominant AEs is in the range of the experiment observation, $40 - 46$ kHz. Figure~\ref{A3} shows the eigenfunction of $n=4$ and $5$ sub-dominant AEs, $4/8$ and $5/10$ GAEs triggered in the inner plasma region. The growth rate is $5$ times smaller compared to the dominant modes and the width of the eigenfunction is localized between $r/a = 0.25 - 0.35$, reason why these modes are not included in the nonlinear simulations for simplicity.

\section*{Appendix B: EP FLR and EP cyclotron frequency effect on the AE stability}

\begin{figure}[h!]
\centering
\includegraphics[width=0.45 \textwidth]{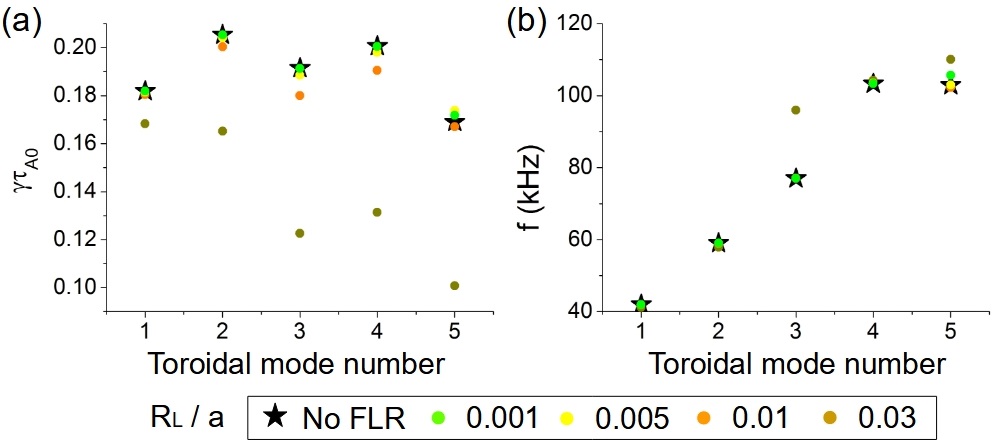}
\caption{Growth rate (a) and frequency (b) of $n=1$ to $5$ AEs in linear simulations without EP FLR damping effects (black stars) and simulations with EP FLR damping effect (circles) for different normalized EP Larmor radius.}\label{B1}
\end{figure}

Figure~\ref{B1} shows the growth rate and frequency of AEs in linear simulations with and without EP FLR damping effects for different Larmor radius. The simulations with normalized Larmor radius $R_{L} / a \leq 0.01$ ($0.006$ m) show growth rates similar to the simulations without EP FLR damping effects, only the simulations with $R_{L} / a = 0.03$ ($0.018$ m) show a significant reduction of the AEs growth rates, particular $n=3$ to $5$ AEs, decreasing by $40 \%$.

\begin{figure}[h!]
\centering
\includegraphics[width=0.45 \textwidth]{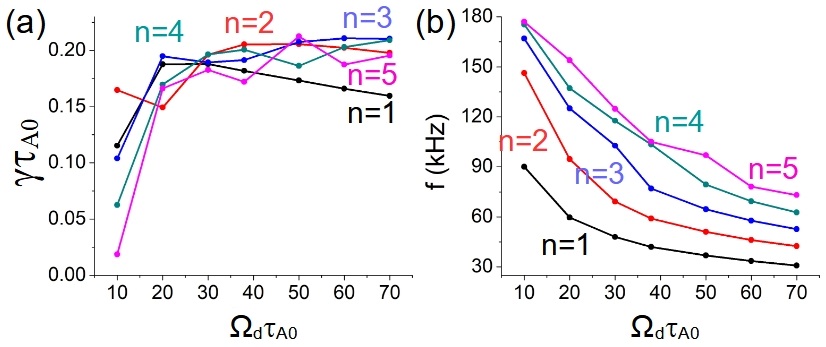}
\caption{Growth rate (a) and frequency (b) of $n=1$ to $5$ AEs in linear simulations for EP with different cyclotron frequencies.}\label{B2}
\end{figure}

Figure~\ref{B2} shows the growth rate and frequency of the AEs in linear simulations for EPs with different normalized cyclotron frequencies ($\Omega_{d} \tau_{A0}$). The simulations show similar growth rates if $\Omega_{d} \tau_{A0} \ge 30$ for all the modes except $n=1$, slightly decreasing as $\Omega_{d} \tau_{A0}$ increases. On the other hand, the AEs frequency decreases as $\Omega_{d} \tau_{A0}$ increases, showing a weaker profile slope for $\Omega_{d} \tau_{A0} \ge 30$. The normalized cyclotron frequency in the linear and nonlinear simulations is $38.78$, located in the parametric range of $\Omega_{d} \tau_{A0}$ values leading to AEs with similar growth rate and frequency. Consequently, a small variation of $\Omega_{d} \tau_{A0}$ should cause small variations in the final result of the simulations.

\hfill \break

\end{document}